\documentclass[fleqn,10pt]{wlscirep_arxiv}

\usepackage{amssymb}
\usepackage{amsfonts} 
\usepackage{textgreek}
\usepackage[squaren,Gray]{SIunits}

\newcommand{\pa}{\partial}

\newcommand{\om}{\omega}

\newcommand{\ga}{\gamma}

\newcommand{\te}{\theta}

\newcommand{\lan}{\langle}
\newcommand{\ran}{\rangle}

\title{Worm-like instability of a vibrated sessile drop}

\author[1]{A. Hemmerle}
\author[1]{G. Froehlicher}
\author[2]{V. Bergeron}
\author[1]{T. Charitat}
\author[1]{J. Farago}
\affil[1]{Universit\'e de Strasbourg, Institut Charles Sadron, CNRS, 23 Rue du Loess, BP 84047, 67034 Strasbourg Cedex 2, France}
\affil[2]{ ENS de Lyon, Laboratoire de Physique, UMR CNRS 5672, 46, all\'ee d'Italie, F69007 Lyon, France}

\begin{abstract}
We study the effects of vertical sinusoidal vibrations on a liquid droplet with a low surface tension (ethanol) deposited on a solid substrate. In a precise range of amplitudes and frequencies, the drop exhibits a dramatic worm-like shape instability with a strong symmetry breaking, comparable to the one observed by Pucci {\it et al.} ({\it Phys. Rev. Lett.}, {\bf 106}, 024503 (2011)) on a vibrated floating lens. However, the geometry of our system is much simpler since it does not involve the oscillation and deformation of a liquid-liquid-air contact line. We show that the Faraday waves appearing on the surface of the droplet control its shape and we draw a systematic phase diagram of the instability. A simple theoretical model allows us to derive a relation between the elongation of the droplet and the amplitude of the Faraday wave, in good agreement with measurements of both quantities.
\end{abstract}
\begin{document}

\flushbottom
\maketitle

\section{Introduction}

Although vibrations of liquid films and drops occur in many practical and technological situations such as coating, spraying, liquid injection and ink-jet printing, the dynamics of the contact line between a liquid, a solid substrate and the surrounding air remains today surprisingly not fully understood \cite{Bonn2009}. The effects of vibrations on liquid drops have been intensively studied \cite{Rodot1979, Brunet2010,vukasinovic(2007),noblin(epje2004),noblin(epjst2009),Noblin2009a}, but a large share of the theoretical and experimental considerations focuses on stationary or quasi-stationary processes where the contact line advances or recedes at slowly varying or constant velocity.  It is, however, well-known that at high accelerations fluid interfaces can develop very specific responses to periodic or quasi-periodic excitations via the parametric amplification of surface waves, the so-called Faraday instability, named after Faraday, who was the first to report their existence \cite{faraday(PhilTrans1831)}.

The Faraday instability has been widely investigated over these last 50 years in the case of large liquid containers \cite{Benjamin1954,Edwards1993,douady(jfluidmech1990)}. Studies of the Faraday instability usually focus on flat fluid surfaces, avoiding the formation of a meniscus on the boundaries, for instance by filling the vessel full to the brim. The presence of a meniscus is known to modify both the onset of the instability and the observed patterns \cite{douady(jfluidmech1990)}. More recently, attention has been focused on Faraday instability in small confined domains with adaptable boundaries and Pucci {\it et al.}  reported the existence of an hydrodynamic instability in an alcohol drop deposited on a cell filled with a viscous  oil, both submitted to vertical oscillations \cite{Pucci(PRL2011),Pucci(JFM2013),Pucci(Nuovo2013),Pucci(Mech2015)}. 

In Pucci {\it et al.}  study, Faraday waves appear on the surface of the floating lens above a certain acceleration threshold but not in the  liquid substrate because of its higher viscosity. Close to this threshold, the system keeps its axial symmetry on average. For higher accelerations, a new threshold is reached, where the system spontaneously breaks the axial symmetry and extends along a random direction. Two different archetypes can be observed depending on whether the drop is wetted or no by the oil. When the oil does not wet the surface of the alcohol drop, the destabilized lens reaches a quasi-stationary state \cite{Pucci(PRL2011), Pucci(JFM2013)}.  The authors developed a model of the elongated shape by considering the interplay between the surface tension at the contact line between the two liquids and air, and the excess pressure due to the Faraday waves. Interestingly, the drop does not reach a steady-state when its surface is wetted by the oil, and keeps extending until it breaks into smaller parts. Because of its large aspect ratio and the standing waves on its surface, this state is thereafter called the ``worm-like state'', following the denomination used by Pucci {\it et al.}  \cite{Pucci(Nuovo2013),Pucci(Mech2015)}

Even in the case of the non-wetted drop, a complete model of their system is rather complex to obtain, in particular because of the presence of three dynamic boundaries (liquid 1-liquid 2, liquid 1-vapor, liquid 2-vapor). Faraday waves do not only appear at the surface of the floating lens, but also at the interface between the lens and the liquid substrate, while a damped wave is generated into this more viscous surrounding liquid by the oscillating lens boundaries. They obtain a good quantitative description of the quasi-stationary shape of the elongated lens, but need to introduce a corrective factor justified by the simplifications made to build their model \cite{Pucci(JFM2013)}.

We describe here the results obtained on a simpler experiment using a drop vibrated on a solid substrate, which shows also a spectacular large symmetry-breaking deformation where a Faraday instability triggers a directed contact line motion under periodical forcing condition. Despite active research on sessile drops under vibrations, the large aspect ratio spreading we observe has, to our knowledge, never been reported on a solid substrate.

\section{Experimental set-up}

\begin{figure}[h!]
 \centering
  \includegraphics[width=90mm]{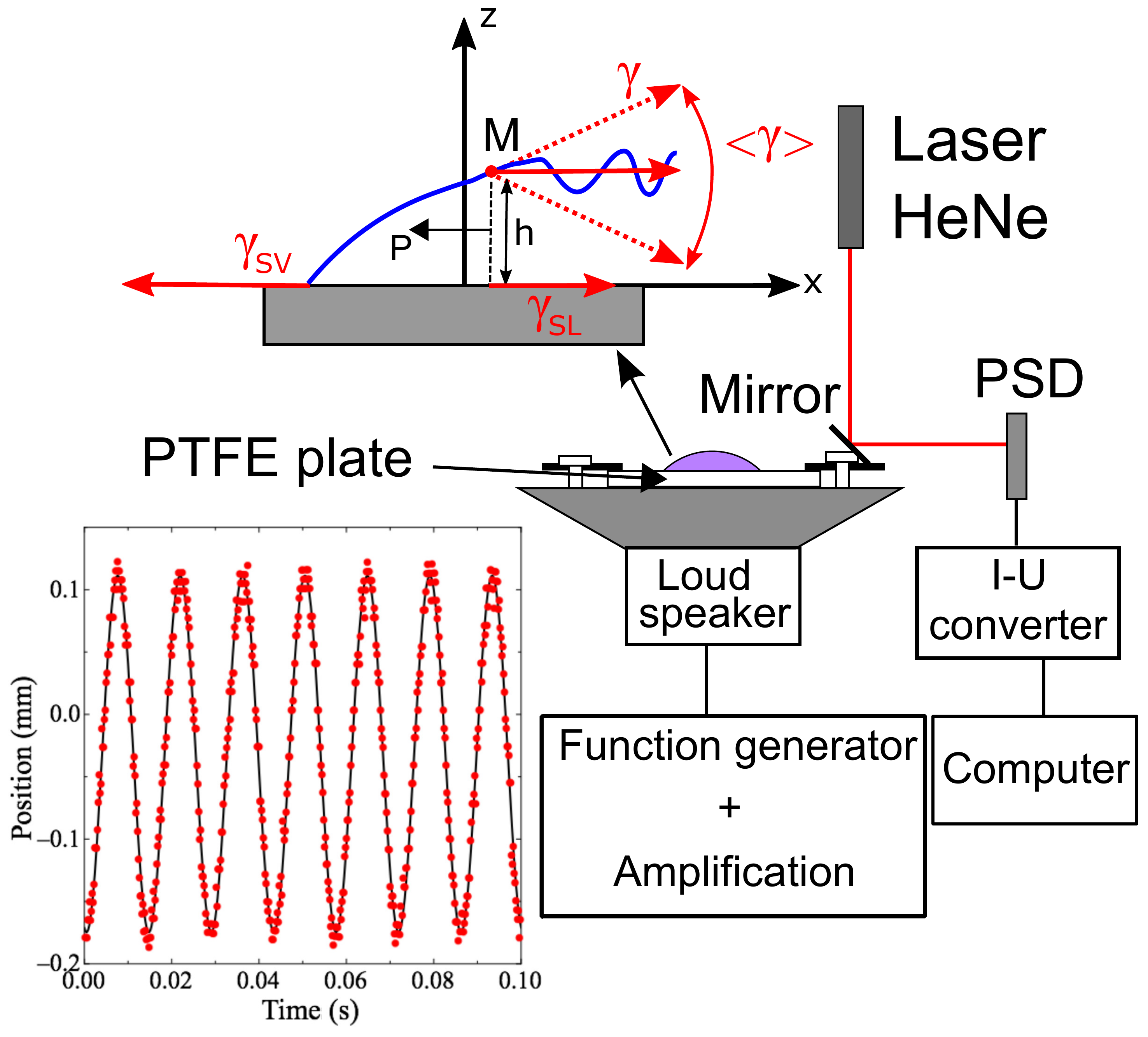} 
\caption{Experimental set-up: the vertical motion of the loudspeaker is measured by a Position-Sensitive Detector (PSD), using the reflection of a laser beam on a mirror fixed to the vibrated substrate. Sinusoidal fits of the data recorded by the PSD give the frequency and the amplitude of the oscillation (inset).} 
\label{figure1} 
\end{figure}

Our experimental set-up (see Fig.\,\ref{figure1}) is similar in many points to the one described by Noblin {\it et al.}  \cite{noblin(epje2004)}. The substrate consists in a plate made of polytetrafluoroethylene (PTFE) fixed on a metal disc. This disc is then bound to the moving part of a loudspeaker, which is connected to a function generator through a power amplifier. We measure the vertical displacement of the substrate by the deflection of a laser beam using a Position-Sensitive Detector (PSD). The system is systematically calibrated using a precision linear stage. This set-up allows us to fix the driving amplitude with a precision of 1$\unit{}{\micro\meter}$. We always work with an excitation signal of the form $U_v = A_v \cos{(\omega_v t)}$ and carefully check that the vibrations of the substrate are purely sinusoidal (see Fig.\,\ref{figure1} inset).

The vibrated drop is imaged with a fast camera. We use the camera from a side-view to measure the amplitude $A_F$ and the time frequency $\omega_k$ of the Faraday waves, and from above to access their wavelength and the shape of the drop.

The drops are made of pure ethanol (surface tension liquid-vapor $\gamma=\unit{22.3}{\milli\newton\per\meter}$, density $\rho=\unit{789}{\kilogram\per\cubic\meter}$, viscosity $\mu=\unit{1.2}{\milli\pascal}$.s), and are placed at the center of the substrate using a micropipette with a fixed volume of \unit{70}{\micro\liter} for this study. The experiments are performed fast enough to avoid a substantial evaporation. The drop is slightly vibrated before any experiment in order to reach its equilibrium state \cite{noblin(epjst2009)}. We measure by side-view pictures an equilibrium contact angle between the PTFE and the ethanol drop of $40^{\circ} \pm 1.5 ^{\circ}$. The height of the drop varies from $1.3$ mm at equilibrium to $620 \pm 30 \unit{}{\micro\meter}$ for the maximal elongation. The wavelength of the Faraday wave is equal to $3.5 \pm 0.2$ mm when the destabilization is reached at 130 Hz. 

\section{Results and discussion}

\begin{figure}[h!] 
\centering
\includegraphics[width=90mm]{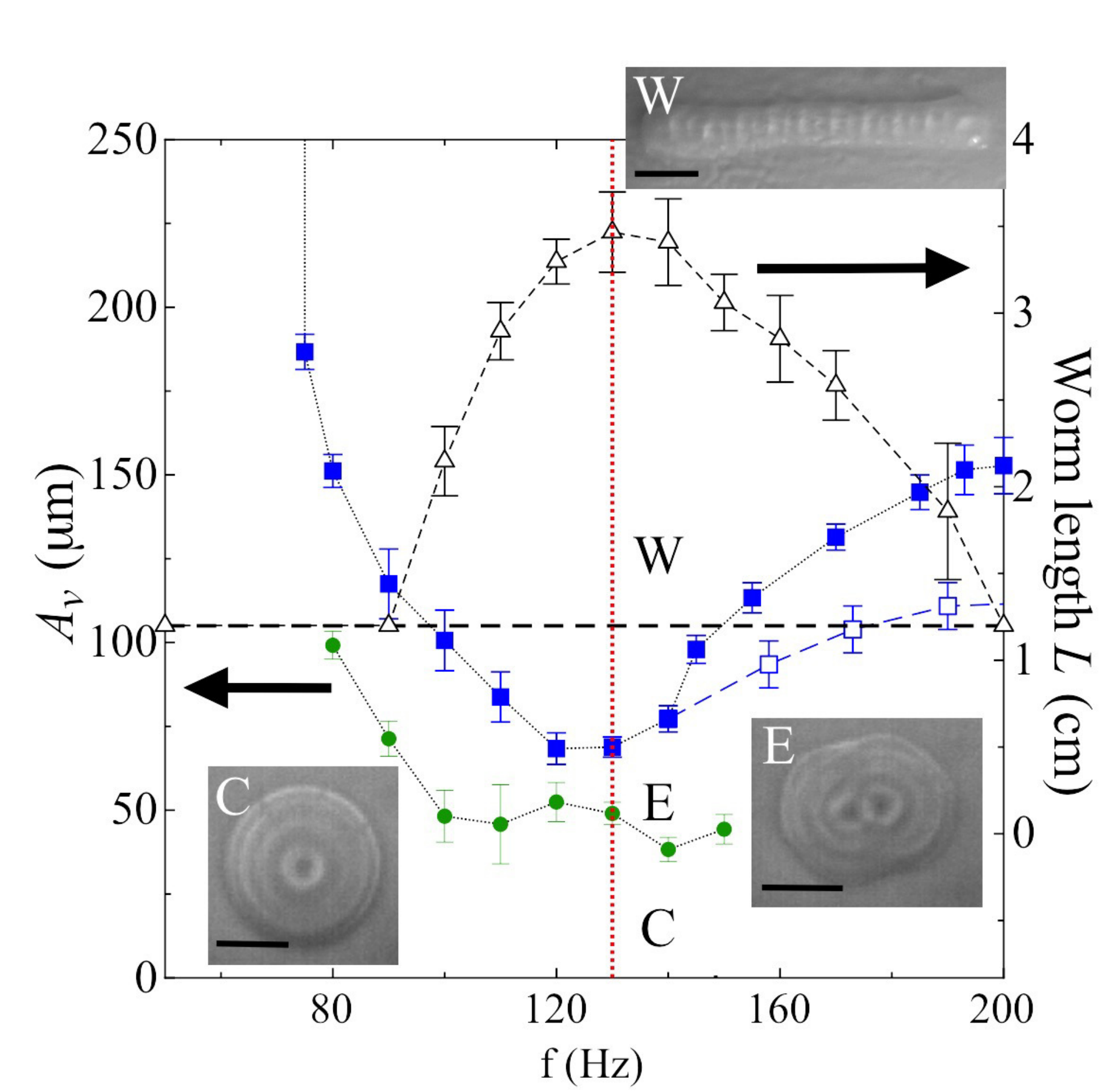} 
\caption{Experimental stability curve of a 70 \textmu L ethanol drop on a PTFE substrate. Filled green circles (\textcolor{green}{$\bullet$}): breakdown of the axisymmetry. Blue squares: transition to the worm-like instability, obtained by keeping the frequency fixed and increasing (filled blue squares (\textcolor{blue}{$\blacksquare$})) or decreasing (open blue squares (\textcolor{blue}{$\square$})) the forcing amplitude $A_v$.  (C): circular state; (E): two-eyes state; (W): worm-like shape. Open black triangles ($\bigtriangleup$): length of the worm measured at a fixed value of $A_v= \unit{105}{\micro\meter}$ for different frequencies (following the horizontal dashed line). Vertical dotted line corresponds to the experiments performed at constant frequency (see Fig.\,\ref{figure3}). The scale bars are \unit{5}{\milli\meter} long.}
\label{figure2}
\end{figure}

We focus first on the different transitions between the drop at rest and the destabilization to the elongated worm-like shape. To draw the phase diagram (Fig.\,\ref{figure2}), we fix the driving frequency and vary slowly the amplitude of oscillation. Error bars represent the statistical dispersion after several measurements of the same value (typically ten), with each time a different drop.

On the lower part of the phase diagram, the drop is axisymmetric and has an average diameter of \unit{12}{\milli\meter} (phase C on Fig.\,\ref{figure2}).  The oscillation of the surface of the drop is harmonic and the contact line remains pinned to the substrate because of the presence of defects \cite{noblin(epjst2009)}. After a threshold amplitude, the drop suddenly loses its circular symmetry, giving rise to a slightly deformed state of two overlapping axisymmetric modes, with subharmonic surface oscillations. We call this configuration the ``two-eyes'' state (E). Interestingly, this state shows strong similarities with the (1,3) mode of destabilization obtained in both solid circular cavities \cite{Ciliberto1985} and floating lenses \cite{Pucci(JFM2013)} submitted to vertical oscillations.
 We observe that the subharmonic response occurs for amplitudes slightly below this threshold, but does not destabilize the drop, probably because of the pinning of the contact line.  The size and shape of the drop are not changing when the forcing amplitude is increased, until the threshold to the worm-like state (W), where the elongation of the drop starts to increase with the amplitude of oscillations.

This last transition is sharp and reversible, {\it i.e.} decreasing the excitation amplitude or changing the frequency can bring  the worm back to the two-eyes state. The position of this transition line on the phase diagram depends on whether the amplitude of oscillation is increased or decreased. This hysteresis is non-negligible only for frequencies higher than \unit{140}{\hertz}. We checked that the direction of elongation is chosen randomly by the system at the transition, showing a real symmetry breaking not induced by the defects of the substrate.

Once the drop forms a worm, we can control its length by changing the frequency (see Fig.\,\ref{figure2}) or the amplitude of oscillation (see Fig.\,\ref{figure3}.A, \textcolor{blue}{$\bullet$}). For a constant driving amplitude, the elongation of the worm is maximal for \unit{130}{\hertz}, which is also the frequency of the minimum amplitude threshold to the worm-like state. The lateral width of the worm $w =5.1 \pm 0.1\unit{}{\milli\meter}$ remains constant and uniform for any elongation.   We denote here fundamental differences between our system and the one of Pucci {\it et al.} 
The worm they obtain when the floating drop is wetted by the surrounding oil (for example with ethanol deposited on silicon oil in Ref.\cite{Pucci(Mech2015)}) keeps also a constant width during the elongation. However, their system is not stable: once the destabilization threshold is reached, the worm extends until breaking into smaller parts. A steady elongated state can be obtained when the drop is not wetted by the oil (isopropanol drop floating on perfluorated oil in Ref.\cite{Pucci(PRL2011)}), and also shows similarities with our system of a drop deposited on a solid substrate. They both exhibit Faraday waves on the surface of the drop, and a strong dependence between the elongation and the forcing amplitude. Nevertheless, Pucci {\it et al.}  observed in this case a width varying with the distance to the tip, and which decreases as the worm elongates (the projected surface remains constant). Volume conservation imposes then that the worm thickness $h$ obtained on the liquid-liquid interface is constant, while $h$ decreases with the elongation in our configuration.

\subsection{Elongation-forcing relation}

A comprehensive theoretical description of the instability would certainly be difficult. A simpler approach consists in bypassing the onset of the worm-like instability (in particular the mechanism by which the contact line wobbles and eventually breaks the circular symmetry), and assuming a well-defined worm-like shape of the drop. Within this regime, a simple theoretical approach is possible to get the ``susceptibility'' of the worm elongation with respect to the amplitude $A_F$ of the Faraday wave. 

The quasi-stationary worm-like shape of the droplet implies that a force budget must exist at each end of the worm, the elongational force manifestly driven by the Faraday waves being equilibrated in a way or another. In Fig.\,\ref{figure1} one of the edge of the worm is depicted,  the Faraday waves are therefore transverse to the figure. The right end of the edge is roughly defined at the beginning of the horizontal oscillating surface; when the worm-like droplet is in a stationary oscillating situation, the edge is chosen in such a way that the point $M$ in the figure is at a node of the surface wave. The forces (expressed per unit length transversal to the figure) exerted on the edge are depicted: the advancing forces are on one hand the solid-vapor surface tension $\ga_{SV}$, and on the other hand the pressure force $\bf F$; the receding forces are the liquid-air and solid-liquid surface tensions $\ga$ and $\ga_{SL}$. At equilibrium, the receding and advancing forces compensate each other, leading to the well-known formula for the equilibrium height of a puddle $h_{\rm eq}=\sqrt{2\ga(1-\cos\te)/\rho g}$, where $\rho$ is the fluid mass density, $g$ the acceleration of gravity and $\te$ the equilibrium value of the liquid-solid and liquid-vapor interface angle \cite{Langmuir1933}. When subjected to a rapid oscillation of the plate, the liquid will develop a Faraday wave far from the edges, whose amplitude $A_F$ is determined mainly by the nonlinear term of the amplitude equation which saturates the linear growth (in case of very shallow water conditions  the liquid depth $h$ also matters). The qualitative effect of this wave is two-fold: firstly, it reduces the average intensity of the tension exerted by the liquid-vapor interface by tilting periodically the vector out of the horizontal (see Fig.\,\ref{figure1}). As a result the averaged vector $\lan\bf\ga\ran$, though still horizontal, is less efficient to pull the edge to the right.
Secondly, an oscillating  flow develops within the fluid below the free surface, which, for a given thickness $h$,  makes also the time-averaged value of the total pressure force $\lan\bf F\ran$ depart from the equilibrium value $|\bf F|=\rho gh^2/2$. 

To evaluate quantitatively these effects, one assumes an irrotational flow described by a potential $\varphi(x,z,t)=A_F\om_k [k\sinh(kh)]^{-1}$\\$\cosh(k[z+h])\sin(kx)\cos(\om_k t)$. From this potential, one gets for the surface deformation $\zeta(x,t)=A_F\sin(kx)\sin(\om_k t)+\xi(x)$, assuming the point $M$ is at $x=0$. The constant deformation $\xi(x)$ is of order $A_F^2$, therefore this expression can be considered as an expansion of $\zeta$ up to the order $A_F^2$, with the proviso that all terms $\propto \sin(2\om_k t)$ are omitted.

The next step of the calculation is the evaluation of the classical Bernoulli relation $P/\rho=-\pa_t\varphi-\frac{v^2}{2}-G(t)z+C$,\\ ($P$ is the gauge pressure) at a point $(x,\zeta(x,t))$ of the surface, with the boundary condition $P(x,\zeta(x))=-\ga\pa^2_{xx}\zeta(x)$, and with $G(t)=g-4\om_k^2A_v\cos(2\om_k t+\psi)$ where $A_v\cos(2\om_k t+\psi)$ is the height of the vibrating plate. $A_v$ is the amplitude of the oscillation of the plate, and one assumes for every frequency a Faraday resonance: the mode actually selected oscillates at a frequency $\om_k/(2\pi)$ which is exactly half that of the excitation, with possibly a phase $\psi$. An expansion in terms of trigonometric functions of $\om t$ and $kx$ is obtained from the Bernoulli relation and the expressions of $\varphi$ and $\zeta$, more precisely it gives various terms proportionnal to products of trigonometric functions of $p\om t$ and $q kx$, where $p$ and $q$ are integers. The orthogonality of these products yields the following results: the phase is either $\psi=0$ or $\pi$. The dispersion relation is renormalized by the driving acceleration according to\footnote{This renormalization has not been taken into account in Ref.\cite{Pucci(JFM2013)}}
\begin{equation}
  \om_k^2=\left(\frac{\ga}{\rho}k^3+gk\right)\frac{\tanh(kh)}{1\mp2kA_v\tanh(kh)\cos(\psi)}
 \label{dispersionrelation}
\end{equation}
with a minus sign if $\psi=0$ and a plus sign if $\psi=\pi$. Here $\om_k$ represents the frequency of the wave, and is equal to half the driving frequency $\om_v=2\om_k$. A third relation allows to compute $C=(A_F\om_k\sinh(kh)^{-1})^2/8$ which is important for the computation of the pressure force. 

Defining the system as the molecules located to the left of $M$ at $t=0$, its right boundary $X(z,t)$ oscillates for $t>0$, and is given by $X(z,t)=A_F\sinh(kh)^{-1}\cosh(k[z+h])\sin(\om_k t)$. A careful computation of the average pressure force exerted on the frontier, up to the second order in $(A_F\om_k)$ (the leading factor of $\varphi$), gives, in accordance with \cite{Longuet(DeepSea1964)},
\begin{eqnarray}
  \lan F_x\ran&=& \frac{\rho g h_{\rm eq}^2}{2}\Bigg(\left(\frac{h}{h_{\rm eq}}\right)^2\frac{(A_Fk)^2}{8(1-\cos\te_E)}\bigg(1+
  \frac{2kh}{\sinh(kh)\cosh(kh)}\bigg)\Bigg).
\end{eqnarray}
 The stationary condition of the edge implies then:
\begin{eqnarray}
 1-\left(\frac{h}{h_{\rm eq}}\right)^2 = \frac{(A_F k)^2}{8(1-\cos\te_E)(1\mp 2kA_v\tanh(kh))}\bigg(3+
\frac{2kh}{\sinh{\left(kh\right)}\cosh{\left(kh\right)}}\bigg).
\label{masterequation}
\end{eqnarray}

This formula is an implicit relation between the stationary height of the worm $h$ and the amplitude $A_F$ of the Faraday wave. If the worm stays in a situation for which $kh\gg 1$, $\tanh(kh)\sim1$, and eq. \ref{dispersionrelation} becomes the dispersion relation for gravity-capillary waves. Last term of eq. \ref{masterequation} would vanish and all the dependence with respect to $h$ would be explicit. The experiments reported here are in conditions of shallow flow ($kh \lesssim 1$), therefore the dispersion relation is in this case explicitly dependent on the height.

\begin{figure}[h!] 
\centering
\includegraphics[width=80mm]{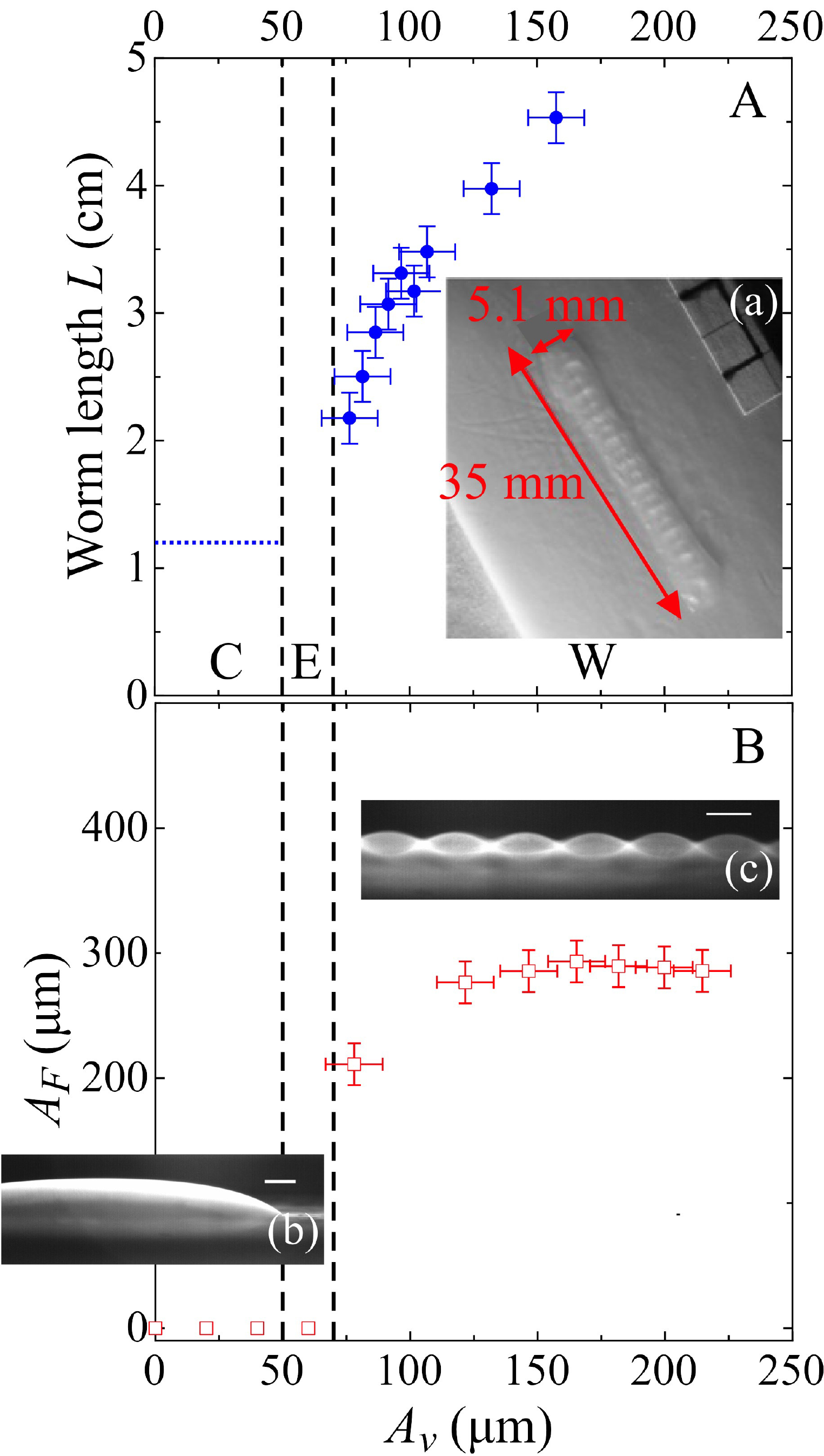} 
\caption{Results of experiments at constant frequency ($f=\unit{130}{\hertz}$) following the vertical dotted line of Fig.\,\ref{figure2}. Top: filled blue circles (\textcolor{blue}{$\bullet$}) show worm length {\it vs} amplitude of vibration $A_v$. The blue dotted line corresponds to the circular drop (C) diameter. Bottom: open red squares (\textcolor{red}{$\square$}) show amplitude of the Faraday waves $A_F$ {\it vs.} amplitude of vibration $A_v$. The dashed lines correspond to the thresholds between circular (C), two-eyes (E) and worm-like shapes. Pictures correspond to (a) top-view in (W) state, (b) side view in (C) state and (c) side-view in (W) state. The scale bars are \unit{1}{\milli\meter} long.}
\label{figure3}
\end{figure}

In principle, it could be possible to express the elongation of the worm in terms of the frequency $\omega_v$ and amplitude $A_v$ of the vibrating plate, but the theoretical expression of the nonlinearly saturated amplitude of Faraday waves is still the object of controversy \cite{douady(jfluidmech1990),Chen1999}.

We measure then directly the amplitude of the Faraday waves $A_F$ with a side-view of the worm for different driving amplitudes at a fixed frequency of \unit{130}{\hertz} (open red squares (\textcolor{red}{$\square$}) on Fig.\,\ref{figure3}.B). The amplitudes measured are ranging from $A_F=\left(210 \pm 20\right) \unit{}{\micro\meter}$ at the onset of the instability to $A_F=\left(290\pm 10\right)\unit{}{\micro\meter}$ for highly elongated worms.
For our range of accelerations (5.2-10.7$g$), the behaviour of $A_F$ is far from the one observed by Pucci {\it et al.}, where they obtain the relation $A_F^2 \propto A_v$ at a fixed frequency.  Here the amplitude of the Faraday waves saturates when approaching half the thickness of the elongated worm $h=620 \pm 30 \unit{}{\micro\meter}$.

\begin{figure}[h!] 
\centering
\includegraphics[width=80mm]{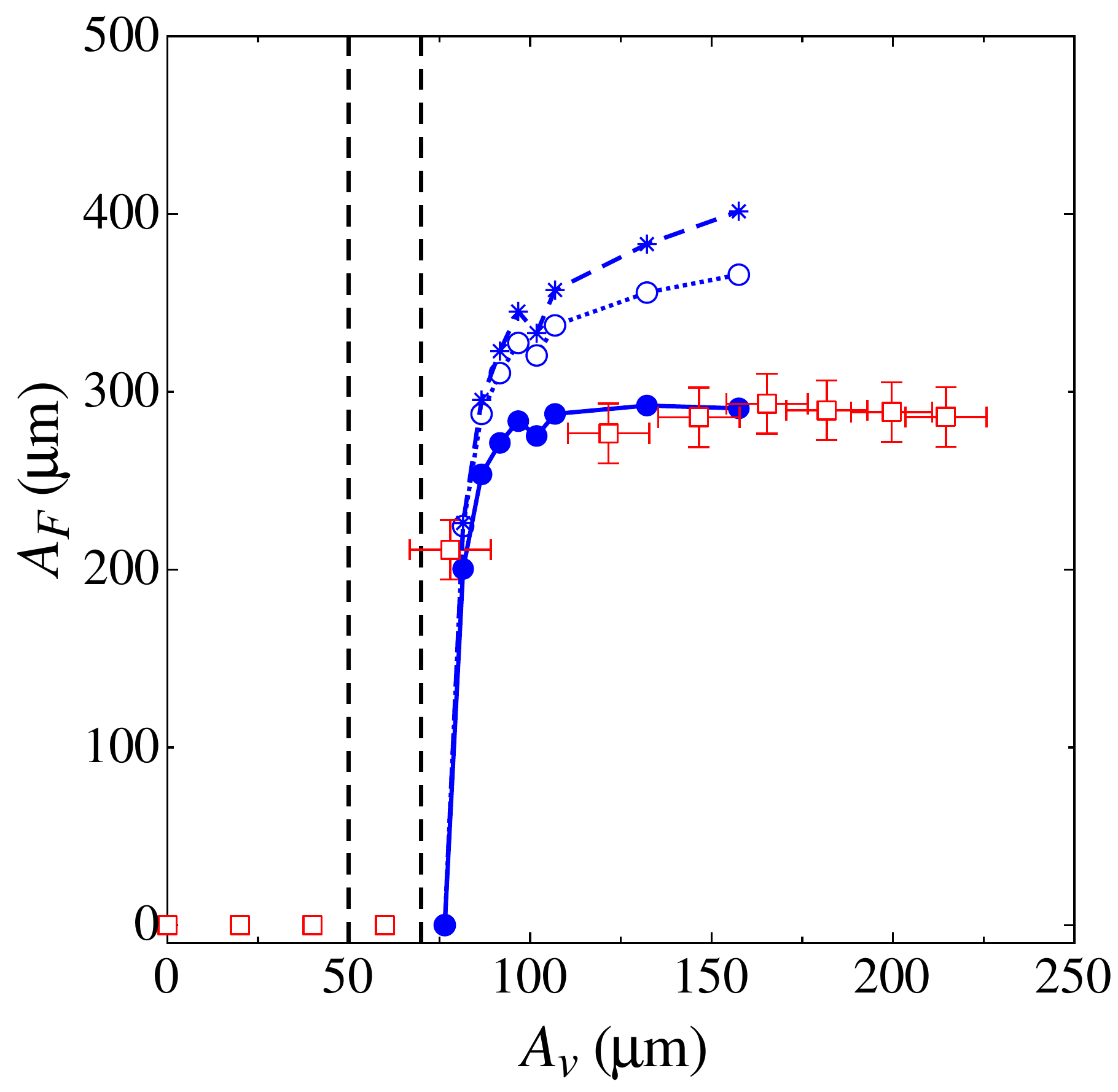} 
\caption{Amplitude of the Faraday waves as a function of the driving amplitude for a fixed frequency of \unit{130}{\hertz}. Open red square (\textcolor{red}{$\square$}): amplitude of the Faraday waves directly measured using a side-view camera. Deduced values of $A_F$ from the measurements of the length of the worm using the usual dispersion relation (blue star (\textcolor{blue}{$\star$}), dashed line) and the renormalized one, eq. \ref{masterequation}, with respectively $\Psi=\pi$ (open blue circles (\textcolor{blue}{$\circ$}), dotted line) or $\Psi=0$ (filled blue circles (\textcolor{blue}{$\bullet$}), solid line). Lines are guides for the eye.
} 
\label{figure4} 
\end{figure}

To compare these direct measures of $A_F$ to the values obtained by solving eq. \ref{masterequation}, we need to determine the thickness of the worm $h$. It is experimentally easier to access its length and then, its width and volume being constant throughout the process, $h/h_{eq}\simeq L/L_0$ with $L$ the measured length of the elongated worm at the driving amplitude $A_v$ and $L_0$ the length of the worm taken at the onset of the instability. Using a phase shift $\psi=0$ we obtain a quantitative agreement between our data and the numerical resolution of eq. \ref{masterequation} (full circles on Fig.\,\ref{figure4}), whereas the other choice $\psi=\pi$ is clearly much less convincing. It demonstrates the validity of our approach and moreover shows that the phases of the vibrating plate and the Faraday wave are locked in phase quadrature.

\section{Conclusion}

We studied experimentally the destabilization of a vibrated sessile droplet in a simple geometry. A steady state is observed, consisting in an elongated worm like state, comparable in shape to the non-steady state describe by Pucci {\it et al.}\cite{Pucci(Nuovo2013),Pucci(Mech2015)}. The apparition of Faraday waves on the surface of the droplet triggers its elongation as shown by our measurements and the theoretical model. The length of the worm saturates with the forcing amplitude because of both its thinning and the increasing amplitude of the Faraday wave. When the forcing amplitude decreases, the radiation pressure decreases too which brings the drop back to a more axisymmetrical state. It is interesting to note that the addition of only a few percents of water to the ethanol strongly shifts the threshold of the instability to much higher driving amplitudes, and even prevents its apparition for volume percentage of water higher than $10\%$. Our experimental configuration is simpler than those of Pucci {\it et al.}\cite{Pucci(JFM2013)}, and a simple model allowed us to account quantitatively for the observed relation between the worm extension and the forcing amplitude in the regime where the worm shape is formed. In particular, we clearly demonstrate that taking into account the renormalization of the dispersion relation by the driving forcing is needed. It remains a substantial challenge to describe theoretically the scenario by which the stationary Faraday waves break the circular symmetry, and why in particular the intermediate two-eyes stage is stable for a range of vibration amplitudes. Finally, an interesting perspective for this work could be an investigation of the response of the vibrating droplet of ethanol in the presence of a microstructured surface, because a low symmetry of the substrate will compete with the symmetry breaking of the droplet and lead potentially to unexpected dynamical stationary states.

\section{Acknowledgments}

We wish to thank Martin Lecourt and Simon Gross for their participation to the experiments.
\newcommand\EatDot[1]{}
 \def\urlprefix{}
 \def\url#1{\EatDot}

\bibliography{bibliogoutte_arxiv}

\begin{thebibliography}{10}
\expandafter\ifx\csname url\endcsname\relax
  \def\url#1{\texttt{#1}}\fi
\expandafter\ifx\csname urlprefix\endcsname\relax\def\urlprefix{URL }\fi
\providecommand{\bibinfo}[2]{#2}
\providecommand{\eprint}[2][]{\url{#2}}

\bibitem{Bonn2009}
\bibinfo{author}{Bonn, D.}, \bibinfo{author}{Eggers, J.},
  \bibinfo{author}{Indekeu, J.}, \bibinfo{author}{Meunier, J.} \&
  \bibinfo{author}{Rolley, E.}
\newblock \bibinfo{title}{Wetting and spreading}.
\newblock \emph{\bibinfo{journal}{Rev. Mod. Phys.}}
  \textbf{\bibinfo{volume}{81}}, \bibinfo{pages}{739--805}
  (\bibinfo{year}{2009}).
\newblock \urlprefix\url{http://link.aps.org/doi/10.1103/RevModPhys.81.739}.

\bibitem{Rodot1979}
\bibinfo{author}{Rodot, H.}, \bibinfo{author}{Bisch, C.} \&
  \bibinfo{author}{Lasek, A.}
\newblock \bibinfo{title}{Zero-gravity simulation of liquids in contact with a
  solid surface}.
\newblock \emph{\bibinfo{journal}{Acta Astronaut.}}
  \textbf{\bibinfo{volume}{6}}, \bibinfo{pages}{1083 -- 1092}
  (\bibinfo{year}{1979}).
\newblock
  \urlprefix\url{http://www.sciencedirect.com/science/article/pii/0094576579900572}.

\bibitem{Brunet2010}
\bibinfo{author}{Brunet, P.}, \bibinfo{author}{Baudoin, M.},
  \bibinfo{author}{Matar, O.~B.} \& \bibinfo{author}{Zoueshtiagh, F.}
\newblock \bibinfo{title}{Droplet displacements and oscillations induced by
  ultrasonic surface acoustic waves: A quantitative study}.
\newblock \emph{\bibinfo{journal}{Phys. Rev. E}} \textbf{\bibinfo{volume}{81}},
  \bibinfo{pages}{036315} (\bibinfo{year}{2010}).
\newblock \urlprefix\url{http://link.aps.org/doi/10.1103/PhysRevE.81.036315}.

\bibitem{vukasinovic(2007)}
\bibinfo{author}{Vukasinovic, B.}, \bibinfo{author}{Smith, M.~K.} \&
  \bibinfo{author}{Gletzer, A.}
\newblock \bibinfo{title}{Dynamics of a sessile drop in forced vibration}.
\newblock \emph{\bibinfo{journal}{J. Fluid Mech.}}
  \textbf{\bibinfo{volume}{587}}, \bibinfo{pages}{395--423}
  (\bibinfo{year}{2007}).
\newblock \urlprefix\url{http://dx.doi.org/10.1017/S0022112007007379}.

\bibitem{noblin(epje2004)}
\bibinfo{author}{Noblin, X.}, \bibinfo{author}{Buguin, A.} \&
  \bibinfo{author}{Brochard-Wyart, F.}
\newblock \bibinfo{title}{Vibrated sessile drops: Transition between pinned and
  mobile contact line oscillations}.
\newblock \emph{\bibinfo{journal}{Eur. Phys. J. E}}
  \textbf{\bibinfo{volume}{14}}, \bibinfo{pages}{395--404}
  (\bibinfo{year}{2004}).
\newblock \urlprefix\url{http://dx.doi.org/10.1140/epje/i2004-10021-5}.

\bibitem{noblin(epjst2009)}
\bibinfo{author}{Noblin, X.}, \bibinfo{author}{Buguin, A.} \&
  \bibinfo{author}{Brochard-Wyart, F.}
\newblock \bibinfo{title}{Vibrations of sessile drops}.
\newblock \emph{\bibinfo{journal}{Eur. Phys. J. ST}}
  \textbf{\bibinfo{volume}{166}}, \bibinfo{pages}{7--10}
  (\bibinfo{year}{2009}).
\newblock \urlprefix\url{http://www.springerlink.com/content/ug6m155hg4x4v2p0}.

\bibitem{Noblin2009a}
\bibinfo{author}{Noblin, X.}, \bibinfo{author}{Kofman, R.} \&
  \bibinfo{author}{Celestini, F.}
\newblock \bibinfo{title}{Ratchetlike motion of a shaken drop}.
\newblock \emph{\bibinfo{journal}{Phys. Rev. Lett.}}
  \textbf{\bibinfo{volume}{102}}, \bibinfo{pages}{194504}
  (\bibinfo{year}{2009}).
\newblock
  \urlprefix\url{http://link.aps.org/doi/10.1103/PhysRevLett.102.194504}.

\bibitem{faraday(PhilTrans1831)}
\bibinfo{author}{Faraday, M.}
\newblock \bibinfo{title}{On the forms and states assumed by fluids in contact
  with vibrating elastic surfaces.}
\newblock \emph{\bibinfo{journal}{Phil. Trans. R. Soc. London}}
  \textbf{\bibinfo{volume}{52}}, \bibinfo{pages}{319--340}
  (\bibinfo{year}{1831}).

\bibitem{Benjamin1954}
\bibinfo{author}{Benjamin, T.~B.} \& \bibinfo{author}{Ursell, F.}
\newblock \bibinfo{title}{The stability of the plane free surface of a liquid
  in vertical periodic motion}.
\newblock \emph{\bibinfo{journal}{Proc. R. Soc. London Ser. A}}
  \textbf{\bibinfo{volume}{225}}, \bibinfo{pages}{505--515}
  (\bibinfo{year}{1954}).

\bibitem{Edwards1993}
\bibinfo{author}{Edwards, W.~S.} \& \bibinfo{author}{Fauve, S.}
\newblock \bibinfo{title}{Parametrically excited quasicrystalline surface
  waves}.
\newblock \emph{\bibinfo{journal}{Phys. Rev. E}} \textbf{\bibinfo{volume}{47}},
  \bibinfo{pages}{R788--R791} (\bibinfo{year}{1993}).

\bibitem{douady(jfluidmech1990)}
\bibinfo{author}{Douady, S.}
\newblock \bibinfo{title}{Experimental study of faraday instability}.
\newblock \emph{\bibinfo{journal}{J. Fluid Mech.}}
  \textbf{\bibinfo{volume}{221}}, \bibinfo{pages}{383--409}
  (\bibinfo{year}{1990}).

\bibitem{Pucci(PRL2011)}
\bibinfo{author}{Pucci, G.}, \bibinfo{author}{Fort, E.},
  \bibinfo{author}{Ben~Amar, M.} \& \bibinfo{author}{Couder, Y.}
\newblock \bibinfo{title}{Mutual adaptation of a faraday instability pattern
  with its flexible boundaries in floating fluid drops}.
\newblock \emph{\bibinfo{journal}{Phys. Rev. Lett.}}
  \textbf{\bibinfo{volume}{106}}, \bibinfo{pages}{024503}
  (\bibinfo{year}{2011}).
\newblock
  \urlprefix\url{http://link.aps.org/doi/10.1103/PhysRevLett.106.024503}.

\bibitem{Pucci(JFM2013)}
\bibinfo{author}{Pucci, G.}, \bibinfo{author}{Ben~Amar, M.} \&
  \bibinfo{author}{Couder, Y.}
\newblock \bibinfo{title}{Faraday instability in floating liquid lenses: the
  spontaneous mutual adaptation due to radiation pressure}.
\newblock \emph{\bibinfo{journal}{J. Fluid Mech.}}
  \textbf{\bibinfo{volume}{725}}, \bibinfo{pages}{402--427}
  (\bibinfo{year}{2013}).
\newblock
  \urlprefix\url{http://journals.cambridge.org/article_S0022112013001663}.

\bibitem{Pucci(Nuovo2013)}
\bibinfo{author}{Pucci, G.}
\newblock \bibinfo{title}{Faraday instability in deformable domains}.
\newblock \emph{\bibinfo{journal}{Nuovo Cimento C}}
  \textbf{\bibinfo{volume}{4}}, \bibinfo{pages}{61--70} (\bibinfo{year}{2013}).

\bibitem{Pucci(Mech2015)}
\bibinfo{author}{Pucci, G.}
\newblock \bibinfo{title}{Faraday instability in floating drops out of
  equilibrium: Motion and self-propulsion from wave radiation stress}.
\newblock \emph{\bibinfo{journal}{Int. J. Non-Linear Mech.}}
  \textbf{\bibinfo{volume}{75}}, \bibinfo{pages}{107--114}
  (\bibinfo{year}{2015}).
\newblock
  \urlprefix\url{http://www.sciencedirect.com/science/article/pii/S0020746215000542}.

\bibitem{Ciliberto1985}
\bibinfo{author}{Ciliberto, S.} \& \bibinfo{author}{Gollub, J.}
\newblock \bibinfo{title}{Chaotic mode competition in parametrically forced
  surface waves}.
\newblock \emph{\bibinfo{journal}{J. Fluid Mech.}}
  \textbf{\bibinfo{volume}{158}}, \bibinfo{pages}{381--398}
  (\bibinfo{year}{1985}).
\newblock
  \urlprefix\url{http://journals.cambridge.org/article_S0022112085002701}.

\bibitem{Langmuir1933}
\bibinfo{author}{Langmuir, I.}
\newblock \bibinfo{title}{Oil lenses on water and the nature of monomolecular
  expanded films}.
\newblock \emph{\bibinfo{journal}{J. Chem. Phys.}}
  \textbf{\bibinfo{volume}{1}}, \bibinfo{pages}{756--776}
  (\bibinfo{year}{1933}).
\newblock
  \urlprefix\url{http://scitation.aip.org/content/aip/journal/jcp/1/11/10.1063/1.1749243}.

\bibitem{Longuet(DeepSea1964)}
\bibinfo{author}{Longuet-Higgins, M.} \& \bibinfo{author}{Stewart, R.}
\newblock \bibinfo{title}{Radiation stresses in water waves; a physical
  discussion, with applications}.
\newblock \emph{\bibinfo{journal}{Deep-Sea Res. Oceanogr. Abst.}}
  \textbf{\bibinfo{volume}{11}}, \bibinfo{pages}{529 -- 562}
  (\bibinfo{year}{1964}).
\newblock
  \urlprefix\url{http://www.sciencedirect.com/science/article/pii/0011747164900014}.

\bibitem{Chen1999}
\bibinfo{author}{Chen, P.} \& \bibinfo{author}{Vi\~nals, J.}
\newblock \bibinfo{title}{Amplitude equation and pattern selection in faraday
  waves}.
\newblock \emph{\bibinfo{journal}{Phys. Rev. E}} \textbf{\bibinfo{volume}{60}},
  \bibinfo{pages}{559--570} (\bibinfo{year}{1999}).
\newblock \urlprefix\url{http://link.aps.org/doi/10.1103/PhysRevE.60.559}.

\end{thebibliography}


\end{document}